\documentclass[%
 reprint,
 amsmath,amssymb,
 aps,
 prb,
]{revtex4-2}

\usepackage{amsmath} 
\usepackage{mathrsfs}%
\usepackage{graphicx}% Include figure files
\usepackage{dcolumn}% Align table columns on decimal point
\usepackage{bm}% bold math

\begin{document}

\title{Theory of Dielectric Behavior in Composites}

\author{Lifeng Hao}
\email{hlf@hit.edu.cn}
\author{Fan Li}%
\author{Yongqi Li}%
\author{Siyong Wang}%
\author{Xiaodong He}%
 \email{hexd@hit.edu.cn}
\affiliation{%
 National Key Laboratory of Science and Technology on Advanced Composites in Special Environments\\
 Harbin Institute of Technology, China
}%

\date{\today}

\begin{abstract}
{While the properties of materials at microscopic scales are well described by fundamental quantum mechanical equations and electronic structure theories, the emergent behavior of mesoscopic or macroscopic composites is no longer governed solely by quantum effects. Instead, such systems are dominated by complex heterogeneous architectures and macroscopic interactions, presenting a classical many-body problem with unique complexities that remain less systematically understood than their quantum counterparts. In this work, we develop an operator-based theoretical framework to characterize these systems, using composite dielectric behavior as a paradigmatic example. By integrating effective medium theory with electromagnetic simulation techniques, we construct an operator that rigorously expresses the effective permittivity tensor as an exact functional. Global and local structure-property relationships can be established by analyzing the operator\textquoteright s structure through symmetric singular value decomposition and block operator matrix analysis, respectively. This framework bridges the gap between microscopic physics and macroscopic material behavior, offering a powerful approach for understanding diverse material properties and guiding the rational design of novel functional composites.}
\end{abstract}

\maketitle

\section{Introduction}
The dielectric response of heterogeneous materials has been a cornerstone of condensed matter physics since the pioneering work of Faraday  \cite{Faraday1857} and Maxwell \cite{Maxwell1881}. While traditional theoretical approaches---effective medium theories (EMTs) \cite{maxwell1904colours, Stroud1975, Choy2015}, bounding methods \cite{Hashin1962}, percolation models \cite{Sahimi1994}, and computational simulations \cite{Myroshnychenko2005, Ghosh1994}---have advanced our understanding, recent breakthroughs in composite fabrication demand a paradigm shift. Modern materials now allow precise control over micro- and nano-scale inclusions, turning the focus from property prediction to uncovering microstructure-property relationships. Key questions remain: How do collective heterogeneities govern macroscopic dielectric response? And how do the size, shape, and spatial arrangement of inclusions contribute to emergent behavior?
A unified theoretical framework for these correlations remains elusive, hindered by two fundamental challenges \cite{Brosseau2006}: (1) the many-body nature of electromagnetic interactions, where local polarization couples to global field distributions, and (2) the interplay between these interactions and the geometric complexity of modern composites. This problem reflects a broader gap in the physics of non-Hermitian systems \cite{Ashida2020}. Unlike quantum systems governed by Hermitian operators (e.g., Schr{\"o}dinger\textquoteright s equation), dielectric response arises from non-Hermitian Helmholtz operators \cite{Choy2015} derived from Maxwell\textquoteright s equations, lacking a rigorous analytical framework. Similar challenges extend to other dissipative systems---thermal, mechanical, and optical---where heterogeneity and collective interactions resist reductionist modeling.

Here, we propose a theoretical framework to bridge this gap. First, we construct an operator in $N$-dimensional Hilbert space that encodes all many-body dielectric interactions, enabling a self-consistent derivation of the permittivity tensor. Second, we analyze the global dielectric response using symmetric singular value decomposition (SSVD), resolving the system into orthogonal states that each contribute to the macroscopic permittivity. Finally, we quantify local behavior via a block operator formalism, explicitly determining interaction ranges and establishing precise structure-property relationships. We validate this framework by analyzing a random spherical dispersion system, demonstrating its potential to unravel the physics of complex composites.
\section{Theory}

\subsection{Problem formulation}
We consider a material domain of volume $\mathscr{V}$ embedded in an infinite composite medium. This representative volume element (RVE) is chosen to be: (i) sufficiently large to contain statistically relevant inhomogeneities, yet (ii) small enough to satisfy quasi-static conditions \cite{Choy2015}. The dielectric response is characterized by a position-dependent permittivity tensor $\bar{\epsilon}(\mathbf{r})$, which for a linear medium relates the displacement field $\mathbf{D}(\mathbf{r})$ to the local electric field $\mathbf{E}(\mathbf{r})$:
\begin{equation}
\mathbf{D}(\mathbf{r})=\bar{\epsilon}(\mathbf{r})\cdot\mathbf{E}(\mathbf{r}).\label{eq:theory_pf1}
\end{equation}
The total electric field comprises contributions from both the applied field $\mathbf{E}_{0}$ (spatially uniform under quasi-static assumptions) and induced currents $\mathbf{J}(\mathbf{r}\textquoteright )$, as described by the integral solution to Maxwell\textquoteright s equations \cite{Weiglhofer1993}:

\begin{equation}
\mathbf{E}(\mathbf{r})=\mathbf{E}_{0}+\int_{\mathscr{V}}\bar{G}_{\mathrm{ee}}(\mathbf{r},\mathbf{r}\textquoteright )\cdot\mathbf{J}(\mathbf{r}\textquoteright )\,dv\textquoteright ,\label{eq:theory_pf2}
\end{equation}
where $\bar{G}_{\mathrm{ee}}(\mathbf{r},\mathbf{r}\textquoteright )$ represents the dyadic Green\textquoteright s function. In the limit of large volume \cite{Stroud1975}, this reduces to the free-space dyadic Green\textquoteright s function. 

Two key challenges emerge: First, $\bar{\epsilon}(\mathbf{r})$ encodes the composite\textquoteright s microstructure but generally lacks closed-form expression. Second, $\mathbf{E}(\mathbf{r})$ depends non-locally on current distributions throughout the medium (Eq.(\ref{eq:theory_pf1})). The combination of complex $\bar{\epsilon}(\mathbf{r})$ and many-body interactions (Eq.(\ref{eq:theory_pf2})) renders dielectric property analysis fundamentally difficult.

\subsection{The effective permittivity tensor}

The dielectric properties of the composite material can be described
by an effective, position-independent permittivity tensor $\bar{\epsilon}_{\mathrm{eff}}$.
To derive its expression, we adopt an approach based on EMT \cite{Stroud1975}.
We begin by expanding the spatially varying permittivity $\bar{\epsilon}(\mathbf{r})$
around an arbitrary reference permittivity $\bar{\epsilon}_{\mathrm{r}}$,
which will later be determined self-consistently. This expansion is
written as:

\begin{equation}
\bar{\epsilon}(\mathbf{r})=\bar{\epsilon}_{\mathrm{r}}+\bar{\chi}(\mathbf{r}),\label{eq:theory_ee1}
\end{equation}
where $\bar{\chi}(\mathbf{r})$ is the electric susceptibility
relative to $\bar{\epsilon}_{\mathrm{r}}$. The corresponding electric
polarization is given by $\mathbf{P}(\mathbf{r})=\bar{\chi}(\mathbf{r})\cdot\mathbf{E}(\mathbf{r})$,
and for time dependence $\exp(i\omega t)$, the polarization current
is $\mathbf{J}(\mathbf{r})=i\omega\mathbf{P}(\mathbf{r})$.
The effective permittivity tensor is then expressed as:

\begin{equation}
\bar{\epsilon}_{\mathrm{eff}}=\bar{\epsilon}_{\mathrm{r}}+\langle\bar{\chi}(\mathbf{r})\rangle,\label{eq:theory_ee2}
\end{equation}
where $\langle\bar{\chi}(\mathbf{r})\rangle$ denotes the ensemble
average of the susceptibility. This average is defined via the macroscopic
polarization $\mathbf{P}_{\mathrm{ave}}$, which relates to $\mathbf{E}_{0}$
as:

\begin{equation}
\mathbf{P}_{\mathrm{ave}}=\langle\bar{\chi}(\mathbf{r})\rangle\cdot\mathbf{E}_{0}.\label{eq:theory_ee3}
\end{equation}
Here, $\mathbf{P}_{\mathrm{ave}}$ is obtained from the volume average:

\begin{equation}
\mathbf{P}_{\mathrm{ave}}=\frac{1}{\mathscr{V}}\int_{\mathscr{V}}\mathbf{P}(\mathbf{r})\,dv.\label{eq:theory_ee4}
\end{equation}
The reference permittivity $\bar{\epsilon}_{\mathrm{r}}$ is determined
self-consistently by enforcing the condition $\langle\bar{\chi}(\mathbf{r})\rangle=0$,
ensuring that the effective medium properly represents the composite\textquoteright s
macroscopic dielectric response.

In conventional EMT \cite{Stroud1975, Choy2015}, $\langle\bar{\chi}(\mathbf{r})\rangle$
is typically derived analytically by making approximations that neglect
detailed geometrical information. In contrast, we adopt an unconventional
approach by expressing $\langle\bar{\chi}(\mathbf{r})\rangle$
explicitly in terms of the complete geometrical configuration using
the method of moments \cite{Gibson2021}. 
We begin by expanding $\mathbf{P}(\mathbf{r})$ as a sum of $N$ local
basis functions:

\begin{equation}
\mathbf{P}(\mathbf{r})=\sum_{n=1}^{N}e_{n}\mathbf{f}_{n}\mathrm{(\mathbf{r}),}\label{eq:theory_ee5}
\end{equation}
where $\{e_{n}\}$ are unknown coefficients. Defining $\langle\mathbf{f}(\mathbf{r}),\mathbf{g}(\mathbf{r})\rangle\equiv\intop_{\mathscr{V}}\mathbf{f}(\mathbf{r})\cdot\mathbf{g}(\mathbf{r})\:dv$
and testing Eq.(\ref{eq:theory_pf2}) with $\mathbf{f}_{m}\mathrm{(\mathbf{r})}$
via Galerkin\textquoteright s method, we obtain: 

\begin{eqnarray}
\langle\mathbf{E\mathrm{(\mathbf{r})}},\mathbf{f}_{m}\mathrm{(\mathbf{r})}\rangle-\sum_{n=1}^{N}e_{n}\langle i\omega\intop_{\mathscr{V}}\bar{G}_{\mathrm{ee}}(\mathbf{r},\mathbf{r\textquoteright })\cdot\mathbf{\mathbf{f}}_{n}(\mathbf{r\textquoteright })\:dv\textquoteright ,\mathbf{f}_{m}\mathrm{(\mathbf{r})}\rangle \nonumber\\ =\langle\mathbf{E_{\mathrm{0}}},\mathbf{f}_{m}\mathrm{(\mathbf{r})}\rangle.\label{eq:theory_ee6}
\end{eqnarray}
This yields a system of $N$ equations for the $N$ unknowns $\{e_{n}\}$,
which can be written in matrix form as:

\begin{equation}
\left[L_{mn}\right]\left[e_{n}\right]=\left[b_{m}\right],\label{eq:theory_ee7}
\end{equation}
where$\left[L_{mn}\right]$ is an $N\times N$ matrix, and $\left[e_{n}\right]$
and $\left[b_{m}\right]$ are column vectors of length $N$. To render
the matrix dimensionless, we normalize each entry of $\left[L_{mn}\right]$
by the average cell volume $\mathscr{V}_{\mathrm{ave}}=\mathscr{V}/N$:

\begin{eqnarray}
&&L_{mn}= \frac{1}{\mathscr{V}_{\mathrm{ave}}}\left\langle \frac{1}{\chi_{n}}\mathbf{f}_{n}(\mathbf{r}),\mathbf{f}_{m}(\mathbf{r})\right\rangle \nonumber\\
&&-\frac{1}{\mathscr{V}_{\mathrm{ave}}}\left\langle i\omega\int_{\mathscr{V}}\bar{G}_{\mathrm{ee}}(\mathbf{r},\mathbf{r}\textquoteright )\cdot\mathbf{f}_{n}(\mathbf{r}\textquoteright )\,dv\textquoteright ,\mathbf{f}_{m}(\mathbf{r})\right\rangle ,\label{eq:theory_ee8}
\end{eqnarray}
where $\chi_{n}$ denotes the scalar susceptibility associated with
$\mathbf{f}_{n}(\mathbf{r})$, and

\begin{equation}
b_{m}=\frac{1}{\mathscr{V}_{\mathrm{ave}}}\left\langle \mathbf{E}_{0},\mathbf{f}_{m}(\mathbf{r})\right\rangle =\mathbf{E}_{0}\cdot\mathbf{p}_{m},\label{eq:theory_ee9}
\end{equation}
where $\mathbf{p}_{m}$ is the normalized electric polarization for
$\mathbf{f}_{m}(\mathbf{r})$:

\begin{equation}
\mathbf{p}_{m}=\frac{1}{\mathscr{V}_{\mathrm{ave}}}\int_{\mathscr{V}}\mathbf{f}_{m}(\mathbf{r})\,dv.\label{eq:theory_ee10}
\end{equation}
Substituting Eq.(\ref{eq:theory_ee9}) into Eq.(\ref{eq:theory_ee7}),
the coefficients $\left[e_{n}\right]$ are obtained as:

\begin{equation}
\left[e_{n}\right]=\left[L_{mn}\right]^{-1}\left[b_{m}\right]=\left[L_{mn}\right]^{-1}\left[(\mathbf{p}_{m})^{T}\right]\cdotp\mathbf{E}_{0},\label{eq:theory_ee11}
\end{equation}
where $\left[(\mathbf{p}_{m})^{T}\right]$ is an $N\times3$ matrix,
with each row representing the transpose of the polarization vector
$\mathbf{p}_{m}$.

The average polarization $\mathbf{P}_{\mathrm{ave}}$ can be expressed
in matrix form by substituting Eqs.(\ref{eq:theory_ee5}) and (\ref{eq:theory_ee10}) into Eq.(\ref{eq:theory_ee4}), yielding

\begin{equation}
\mathbf{P}_{\mathrm{ave}}=\frac{1}{\mathscr{V}}\intop_{\mathscr{V}}\sum_{n=1}^{N}e_{n}\mathbf{f}_{n}\mathrm{(\mathbf{r})\:}dv=\frac{1}{N}\left[(\mathbf{p}_{n})^{T}\right]^{T}\left[e_{n}\right],\label{eq:theory_ee12}
\end{equation}
where $\left[(\mathbf{p}_{n})^{T}\right]^{T}$ denotes the transpose
of $\left[(\mathbf{p}_{n})^{T}\right]$. Substituting Eq.(\ref{eq:theory_ee11})
into Eq.(\ref{eq:theory_ee12}) and using $\left[(\mathbf{p}_{m})^{T}\right]=\left[(\mathbf{p}_{n})^{T}\right]$,
we obtain

\begin{equation}
\mathbf{P}_{\mathrm{ave}}=\frac{1}{N}\left[(\mathbf{p}_{n})^{T}\right]^{T}\left[L_{mn}\right]^{-1}\left[(\mathbf{p}_{n})^{T}\right]\cdotp\mathbf{E}_{0}.\label{eq:theory_ee13}
\end{equation}
Consequently, $\langle\bar{\chi}(\mathbf{r})\rangle$ can be
expressed as

\begin{equation}
\langle\bar{\chi}(\mathbf{r})\rangle=\frac{1}{N}\left[(\mathbf{p}_{n})^{T}\right]^{T}\left[L_{mn}\right]^{-1}\left[(\mathbf{p}_{n})^{T}\right].\label{eq:theory_ee14}
\end{equation}
Notably, $\left[L_{mn}\right]$ depends on $\bar{\epsilon}_{\mathrm{eff}}$.
Therefore, $\bar{\epsilon}_{\mathrm{eff}}$ must be computed iteratively,
with the $(t+1)$-th estimate given by

\begin{equation}
\left(\bar{\epsilon}_{\mathrm{eff}}\right)_{t+1}=\left(\bar{\epsilon}_{\mathrm{eff}}\right)_{t}+\langle\bar{\chi}(\mathbf{r})\rangle_{t},\label{eq:theory_ee15}
\end{equation}
until $\langle\bar{\chi}(\mathbf{r})\rangle_{t}$ converges to
zero. Crucially, $\langle\bar{\chi}(\mathbf{r})\rangle$ depends
on two key terms: (1) $\left[(\mathbf{p}_{n})^{T}\right]$, which
encodes the full geometric and dielectric properties of the system,
and (2) $\left[L_{mn}\right]$, which captures all pairwise interactions
between basis functions. The combination of these terms provides a
complete description of the composite material, rendering $\bar{\epsilon}_{\mathrm{eff}}$
a rigorous and intrinsic dielectric property that is independent of
the applied field.

This framework extends beyond conventional EMTs, enabling exact predictions
of the permittivity tensor for composites with arbitrarily complex
microstructures. Unlike numerical methods such as finite element method \cite{Myroshnychenko2005}
or boundary integral method \cite{Ghosh1994}, which rely on field simulations, our
approach focuses on dipole interactions rather than direct field calculations.
Moreover, the effective permittivity serves as a boundary condition
rather than a spatial boundary, positioning this method as a material
simulation technique analogous to the Kohn-Sham approach \cite{Kohn1965} in density
functional theory \cite{Hohenberg1964}.

A key advantage of this approach is that it eliminates the need to
predefine the applied field $\mathbf{E}_{0}$, as required in traditional
numerical methods. Instead, the polarization distribution $\mathbf{P}(\mathbf{r})$
for an arbitrary $\mathbf{E}_{0}$ can be obtained by substituting
Eq.(\ref{eq:theory_ee11}) into Eq.(\ref{eq:theory_ee5}), yielding: 

\begin{equation}
\mathbf{P}(\mathbf{r})=\left[\left(\mathbf{f}_{n}\mathrm{(\mathbf{r})}\right)^{T}\right]^{T}\left[L_{mn}\right]^{-1}\left[(\mathbf{p}_{m})^{T}\right]\cdotp\mathbf{E}_{0}.\label{eq:theory_ee16}
\end{equation}
Similarly, the electric field distribution $\mathbf{E}(\mathbf{r})$
is given by: 

\begin{equation}
\mathbf{E}(\mathbf{r})=\left[\frac{1}{\chi_{n}}\left(\mathbf{f}_{n}\mathrm{(\mathbf{r})}\right)^{T}\right]^{T}\left[L_{mn}\right]^{-1}\left[(\mathbf{p}_{m})^{T}\right]\cdotp\mathbf{E}_{0}.\label{eq:theory_ee17}
\end{equation}
This facilitates detailed analysis of the composite\textquoteright s
response under diverse external fields, offering deeper insights into
its electromagnetic behavior.

\subsection{Global analysis via SSVD}

The effective permittivity $\bar{\epsilon}_{\mathrm{eff}}$ is expressed
as a functional of the matrix $\left[L_{mn}\right]$, which acts as
a linear operator in an $N$-dimensional Hilbert space $\mathcal{H}$
spanned by the complete basis $\{\mathbf{f}_{n}\mathrm{(\mathbf{r})}\}_{n=1}^{N}$.
Interpreting $\left[L_{mn}\right]$ as an operator allows the application
of diverse analytical tools to probe the underlying physics of dielectric
response encoded in its structure.

Spectral theory \cite{Kreyszig1991} is particularly powerful in this context, as it decomposes
an operator into simpler components, revealing its fundamental structure
and associated physical mechanisms. While the theory is well-established
for Hermitian systems---characterized by real eigenvalues and orthogonal
eigenfunctions---its extension to non-Hermitian operators presents
ongoing challenges. The Helmholtz operator $\left[L_{mn}\right]$
exemplifies these complexities, exhibiting eigenfunctions that generally
form a bi-orthogonal set and lack normalizability \cite{Choy2015, Ashida2020}.

However, $\left[L_{mn}\right]$ possesses complex symmetry, enabling
its factorization via SSVD (also called Takagi factorization) \cite{Horn2012}:

\begin{equation}
[L_{mn}]=Q\Lambda Q^{T}\label{eq:theory_ga1}
\end{equation}
where $Q$ is a unitary matrix and $\Lambda=\textrm{diag}(\lambda_{1},\lambda_{2}...,\lambda_{N})$
is a nonnegative diagonal matrix containing the singular values of
$\left[L_{mn}\right]$. By substituting Eq.(\ref{eq:theory_ga1}) into
Eq.(\ref{eq:theory_ee14}), the ensemble-averaged susceptibility $\langle\bar{\chi}(\mathbf{r})\rangle$
can be expressed as:

\begin{equation}
\langle\bar{\chi}(\mathbf{r})\rangle=\frac{1}{N}\left[(\mathbf{p}_{n})^{T}\right]^{T}Q^{*}\Lambda^{-1}Q^{H}\left[(\mathbf{p}_{n})^{T}\right],\label{eq:theory_ga2}
\end{equation}
where $Q^{*}$ and $Q^{H}$ denote the complex conjugate and Hermitian
transpose of $Q$, respectively. Representing $Q$ as the column-vector
set $\{q_{n}\}_{n=1}^{N}$, where each $q_{n}$ is a singular vector,
the SSVD transforms the original basis $\{\mathbf{f}_{n}\mathrm{(\mathbf{r})}\}_{n=1}^{N}$
into the orthonormal singular vector basis $\{q_{n}\}_{n=1}^{N}$.
In this new basis, each $q_{n}$ corresponds to an independent global
multipole state spanning the entire RVE. 

This basis transition not only simplifies the analysis but also provides
deeper insight into the dielectric behavior governed by the matrix
$[L_{mn}]$. We define a complex vector associated with the state
$q_{n}$ as: 

\begin{equation}
\mathbf{t}_{n}(q_{n})=\left[(\mathbf{p}_{n})^{T}\right]^{T}q_{n}^{*}.\label{eq:theory_ga3}
\end{equation}
In terms of $\mathbf{t}_{n}(q_{n})$, Eq.(\ref{eq:theory_ga2}) can
be rewritten compactly as: 

\begin{equation}
\langle\bar{\chi}(\mathbf{r})\rangle=\frac{1}{N}\sum_{n=1}^{N}\frac{1}{\lambda_{n}}\mathbf{t}_{n}(q_{n})\mathbf{t}_{n}^{T}(q_{n}),\label{eq:theory_ga4}
\end{equation}
where $\mathbf{t}_{n}(q_{n})\mathbf{t}_{n}^{T}(q_{n})$ is a dyad
corresponding to $q_{n}$. For an arbitrary applied field $\mathbf{E}_{0}$,
this dyad yields a polarization response, $\frac{1}{\lambda_{n}}\left[\mathbf{t}_{n}^{T}(q_{n})\cdot\mathbf{E}_{0}\right]\mathbf{t}_{n}(q_{n})$,
which is the complex vector $\mathbf{t}_{n}(q_{n})$ weighted by the
scalar $\frac{1}{\lambda_{n}}\left[\mathbf{t}_{n}^{T}(q_{n})\cdot\mathbf{E}_{0}\right]$,
independent of all other states.
Substituting Eq.(\ref{eq:theory_ga4}) into Eq.(\ref{eq:theory_ee15})
yields: 

\begin{equation}
\bar{\epsilon}_{\mathrm{eff}}=\frac{1}{N}\sum_{n=1}^{N}\bar{\epsilon}(q_{n}),\label{eq:theory_ga5}
\end{equation}
where 
\begin{equation}
\bar{\epsilon}(q_{n})=\bar{\epsilon}_{\mathrm{eff}}+\frac{1}{\lambda_{n}}\mathbf{t}_{n}(q_{n})\mathbf{t}_{n}^{T}(q_{n}).\label{eq:theory_ga6}
\end{equation}
The SSVD thus decouples the overall dielectric behavior into $N$
independent $q_{n}$ multipole states, where each state contributes
a dyadic term $\frac{1}{N}\left[\bar{\epsilon}_{\mathrm{eff}}+\frac{1}{\lambda_{n}}\mathbf{t}_{n}(q_{n})\mathbf{t}_{n}^{T}(q_{n})\right]$
to $\bar{\epsilon}_{\mathrm{eff}}$. This establishes a key relationship
$\bar{\epsilon}(q)\sim q$, as revealed by Eq.(\ref{eq:theory_ga6}),
analogous to the $E(\mathbf{k})\sim\mathbf{k}$ dispersion \cite{Ashcroft1976} in electronic
band theory. Such a relationship provides a fundamental link between
the dielectric response and the underlying structural features.

From the perspective of statistical physics, the SSVD decomposition
allows us to interpret the RVE as a system of $N$ interacting cells.
Here, the set of unknown coefficients $\{e_{n}\}$ defines a microstate
of the system, corresponding to a point in an $N$-dimensional phase
space. Although this space contains infinitely many microstates, only
a subset are physically accessible due to constraints imposed by the
geometric structure and intercellular interactions. The SSVD explicitly
identifies these accessible states, which are encoded in the matrix
$Q$, with each state occupying a phase space density of $1/N$.
In this framework, the system resembles a microcanonical ensemble \cite{Kardar2007},
where the accessible states are equally probable under fixed macroscopic
constraints. Consequently, the ensemble average $\langle\bar{\chi}(\mathbf{r})\rangle$
represents the statistical mean over all permissible microstates.
This interpretation not only reinforces the physical significance
of the SSVD decomposition but also provides a natural connection to
thermodynamic averaging in many-body systems.

\subsection{Local analysis via block operator matrix}

While the $\bar{\epsilon}(q)\sim q$ relationship provides a global
description of composite dielectric behavior, local analysis is critical
for understanding microscale interactions---particularly the response
of individual inclusions or substructures. To achieve this, we decompose
the Hilbert space $\mathcal{H}$ into $\mathcal{H}=\mathcal{H}_{1}\oplus\mathcal{H}_{2}$,
where $\mathcal{H}_{1}$ represents the substructure of interest (e.g.,
an inclusion), and $\mathcal{H}_{2}$ describes its immediate surroundings,
which are further embedded in a homogeneous effective medium with
permittivity $\bar{\epsilon}_{\mathrm{eff}}$. The subspaces $\mathcal{H}_{1}$
and $\mathcal{H}_{2}$ are spanned by bases $\{\mathbf{f}_{n1}\mathrm{(\mathbf{r})}\}_{n1=1}^{N_{1}}$
and $\{\mathbf{f}_{n2}\mathrm{(\mathbf{r})}\}_{n2=1}^{N_{2}}$ , respectively,
with $N_{1}+N_{2}=N$.

Within this framework, the operator matrix $\left[L_{mn}\right]$
adopts a block structure \cite{Tretter2008}: 

\begin{equation}
\left[L_{mn}\right]=\left[\begin{array}{cc}
A_{11} & A_{12}\\
A_{21} & A_{22}
\end{array}\right],\label{eq:theory_la1}
\end{equation}
where $A_{11}$ and $A_{22}$ act within $\mathcal{H}_{1}$ and $\mathcal{H}_{2}$,
respectively, while $A_{12}$ and $A_{21}$ couple the two subspaces.
Its inverse is given by:

\begin{equation}
\left[L_{mn}\right]^{-1}=\left[\begin{array}{cc}
C_{1}^{-1} & B_{1}\\
B_{1}^{T} & C_{2}^{-1}
\end{array}\right],\label{eq:theory_la2}
\end{equation}
with $C_{1}=A_{11}-A_{12}A_{22}^{-1}A_{21}$ and $C_{2}=A_{22}-A_{21}A_{11}^{-1}A_{12}$
being the Schur complements \cite{Horn2012} of $A_{22}$ and $A_{11}$, and $B_{1}=-A_{11}^{-1}A_{12}C_{2}^{-1}$.
The ensemble-averaged susceptibility $\langle\bar{\chi}(\mathbf{r})\rangle$
then decomposes into four contributions: 

\begin{equation}
\langle\bar{\chi}(\mathbf{r})\rangle=\bar{\chi}_{11}+\bar{\chi}_{12}+\bar{\chi}_{21}+\bar{\chi}_{22},\label{eq:theory_la3}
\end{equation}
where 
\begin{eqnarray*}
\bar{\chi}_{11}=\frac{1}{N}\left[(\mathbf{p}_{n1})^{T}\right]^{T}C_{1}^{-1}\left[(\mathbf{p}_{n1})^{T}\right], \\
\bar{\chi}_{22}=\frac{1}{N}\left[(\mathbf{p}_{n2})^{T}\right]^{T}C_{2}^{-1}\left[(\mathbf{p}_{n2})^{T}\right],
\end{eqnarray*}
arise from the intrinsic polarizabilities of $\mathcal{H}_{1}$ and
$\mathcal{H}_{2}$, while 
\begin{eqnarray*}
\bar{\chi}_{12}=\frac{1}{N}\left[(\mathbf{p}_{n1})^{T}\right]^{T}B_{1}\left[(\mathbf{p}_{n2})^{T}\right], \\
\bar{\chi}_{21}=\frac{1}{N}\left[(\mathbf{p}_{n2})^{T}\right]^{T}B_{1}^{T}\left[(\mathbf{p}_{n1})^{T}\right],
\end{eqnarray*}
quantify their mutual interactions. Here, $\mathbf{p}_{n1}$ and $\mathbf{p}_{n2}$
are the normalized polarization vectors for their respective basis
functions.

The local electric fields induced by an applied field $\mathbf{E}_{0}$
further elucidate the microscale interactions between substructures.
The field in $\mathcal{H}_{2}$ generated by the substructure of interest
($\mathcal{H}_{1}$) is given by

\begin{equation}
\mathbf{E}_{21}(\mathbf{r})=\left[\frac{1}{\chi_{n2}}\left(\mathbf{f}_{n2}\mathrm{(\mathbf{r})}\right)^{T}\right]^{T}B_{1}^{T}\left[(\mathbf{p}_{n1})^{T}\right]\cdotp\mathbf{E}_{0},\label{eq:theory_la4}
\end{equation}
representing the dielectric response of the surrounding medium to
$\mathcal{H}_{1}$. Conversely, the field in $\mathcal{H}_{1}$ induced
by $\mathcal{H}_{2}$,

\begin{equation}
\mathbf{E}_{12}(\mathbf{r})=\left[\frac{1}{\chi_{n1}}\left(\mathbf{f}_{n1}\mathrm{(\mathbf{r})}\right)^{T}\right]^{T}B_{1}\left[(\mathbf{p}_{n2})^{T}\right]\cdotp\mathbf{E}_{0},\label{eq:theory_la5}
\end{equation}
quantifies the environment\textquoteright s back-action on the substructure. These
spatially resolved fields, combined with the interaction susceptibilities
$\bar{\chi}_{12}$ and $\bar{\chi}_{21}$, quantitatively link local
structural features to their dielectric response, enabling targeted
analysis of inclusion-surrounding coupling effects.

\section{Numerical results}

\subsection{Model description}

\begin{figure}
\includegraphics[width=0.45\textwidth]{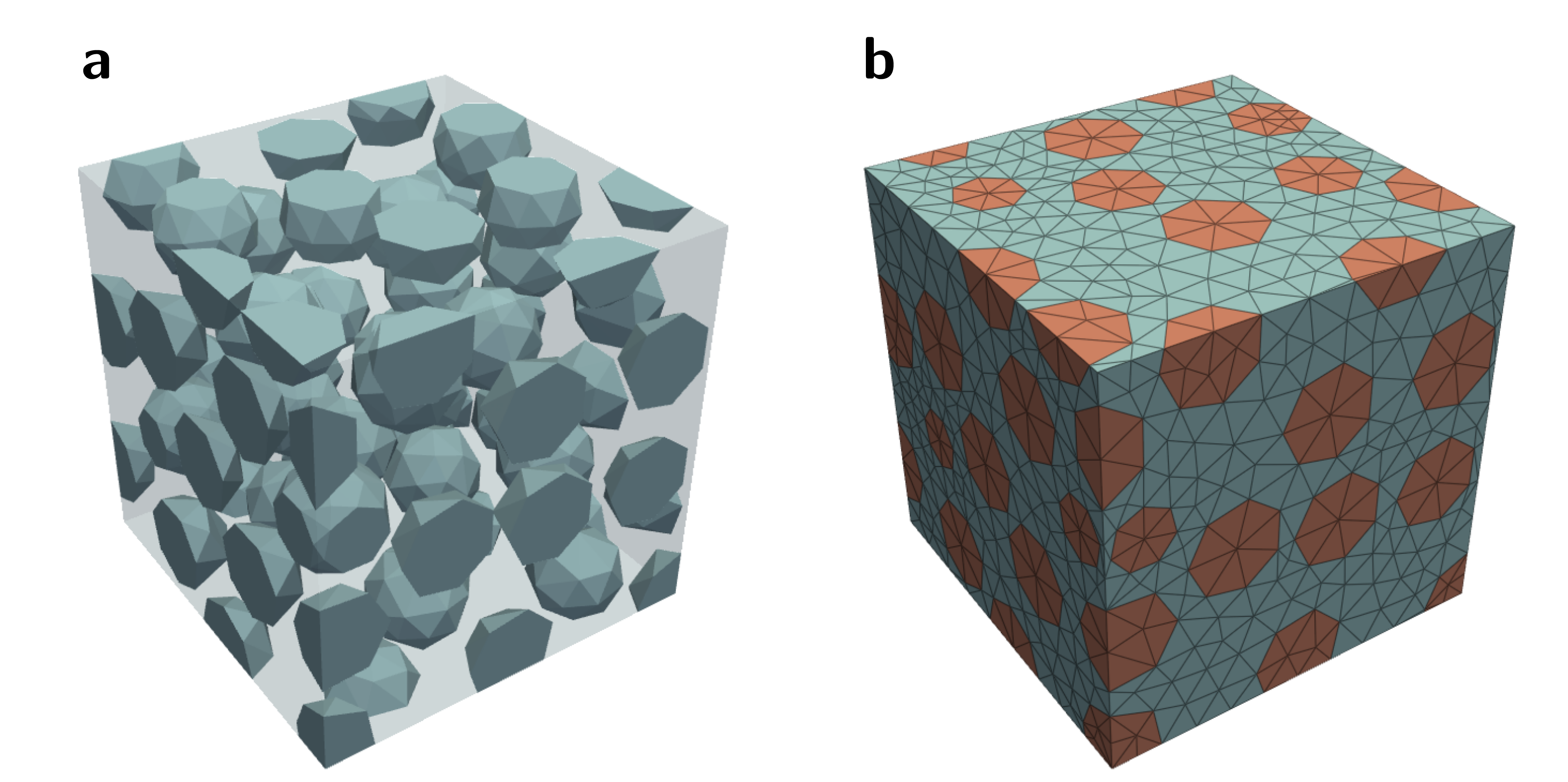}
\caption{\label{fig1} \textbf{Composite model of spherical inclusions in a dielectric host.}
\textbf{a,} Geometrical configuration: 70 identical spherical inclusions
(radius $r_{0}=1.2~\mu$m) randomly distributed within a cubic host
(edge length $10~\mu$m), yielding an inclusion volume fraction of
24.86\%. Inclusions intersecting the boundary are truncated to maintain
a smooth interface. \textbf{b,} Discretized mesh generated using GMSH \cite{Geuzaine2009},
with a target element size of $1.2~\mu$m.}
\end{figure}

To demonstrate the theoretical framework for analyzing composite materials,
we consider an RVE consisting of a continuous dielectric matrix with
randomly dispersed identical spherical inclusions (Fig.\ref{fig1}\textbf{a}). The inclusions
and matrix are modeled as isotropic, homogeneous materials with scalar
permittivities of $\epsilon_{\mathrm{i}}=50.0-5.0i$ (inclusion) and
$\epsilon_{\mathrm{m}}=3.0-0.1i$ (matrix), respectively. The system
is discretized into 9,131 tetrahedral elements (Fig.\ref{fig1}\textbf{b}), generating
36,524 half-Schaubert-Wilton-Glisson (H-SWG) \cite{Zhang2015} basis functions ($N=36,524$).
An applied field frequency of 3 GHz (vacuum wavelength =
10 cm) ensures a quasi-static assumption, as the RVE dimensions
(10 $\mu$m) are orders of magnitude smaller than the wavelength.

Although our theory accommodates complex geometries, we focus on this
idealized system to enable direct comparison with EMT predictions.
The model---featuring isolated inclusions---satisfies EMT\textquoteright s key
assumption that each constituent is surrounded by a homogeneous effective
medium \cite{Choy2015}, thereby facilitating validation against classical EMT results.

\subsection{Calculation of $\bar{\epsilon}_{\mathrm{eff}}$}

The effective permittivity tensor $\bar{\epsilon}_{\mathrm{eff}}$
was computed through five iteration cycles, yielding:
\begin{eqnarray*}
&&\bar{\epsilon}_{\mathrm{eff}}= \\
&&\left(\begin{array}{ccc}
5.894-0.258i & -0.001+0.0i & 0.01-0.001i\\
-0.001+0.0i & 5.893-0.258i & -0.014+0.001i\\
0.01-0.001i & -0.014+0.001i & 5.956-0.262i
\end{array}\right).
\end{eqnarray*}
The tensor is strongly diagonal-dominant, with nearly identical diagonal
elements (average: $5.914-0.259i$), consistent with an isotropic
medium. A slight anisotropy (\ensuremath{\sim}1\% larger magnitude
in the last diagonal component) arises from finite-size effects, while
negligible off-diagonal terms stem from residual anisotropy and minor
numerical artifacts.

For comparison, the Maxwell-Garnett (MG) EMT \cite{maxwell1904colours}  predicts
$\epsilon_{\mathrm{MG}}=5.375-0.213i$, and the volume-averaged local
permittivity gives $14.686-1.318i$. Both our result and the MG value
are significantly smaller than the volume average, indicating strong
depolarization effects \cite{Choy2015}. However, our result exceeds the MG prediction
by \ensuremath{\sim}10\%, a discrepancy attributable to inter-particle
coupling \cite{Sihvola1999}---explicitly included in our model but absent in MG\textquoteright  s
single-particle approximation. The higher permittivity reflects the
expected enhancement from interactions between high-permittivity inclusions.

Our calculations reveal that the dielectric behavior of this system
is governed by two competing factors: (1) depolarization from the
low-permittivity host and (2) field enhancement due to coupling between
high-permittivity inclusions. The dominance of depolarization effects
in $\bar{\epsilon}_{\mathrm{eff}}$ suggests that the former outweighs
the latter. However, key questions remain unresolved:
\begin{itemize}
\item Quantifying depolarization: What is the magnitude of the depolarization
effect?
\item Dominance mechanism: Why does depolarization prevail over coupling?
\item Structural dependence: How do these effects relate to the isolated
dispersion geometry?
\end{itemize}
These questions are critical for uncovering the fundamental physics
of dielectric behavior in such systems but remain unaddressed in prior
studies.

\subsection{Quantifying the depolarization effect}

\begin{figure}
\includegraphics[width=0.45\textwidth]{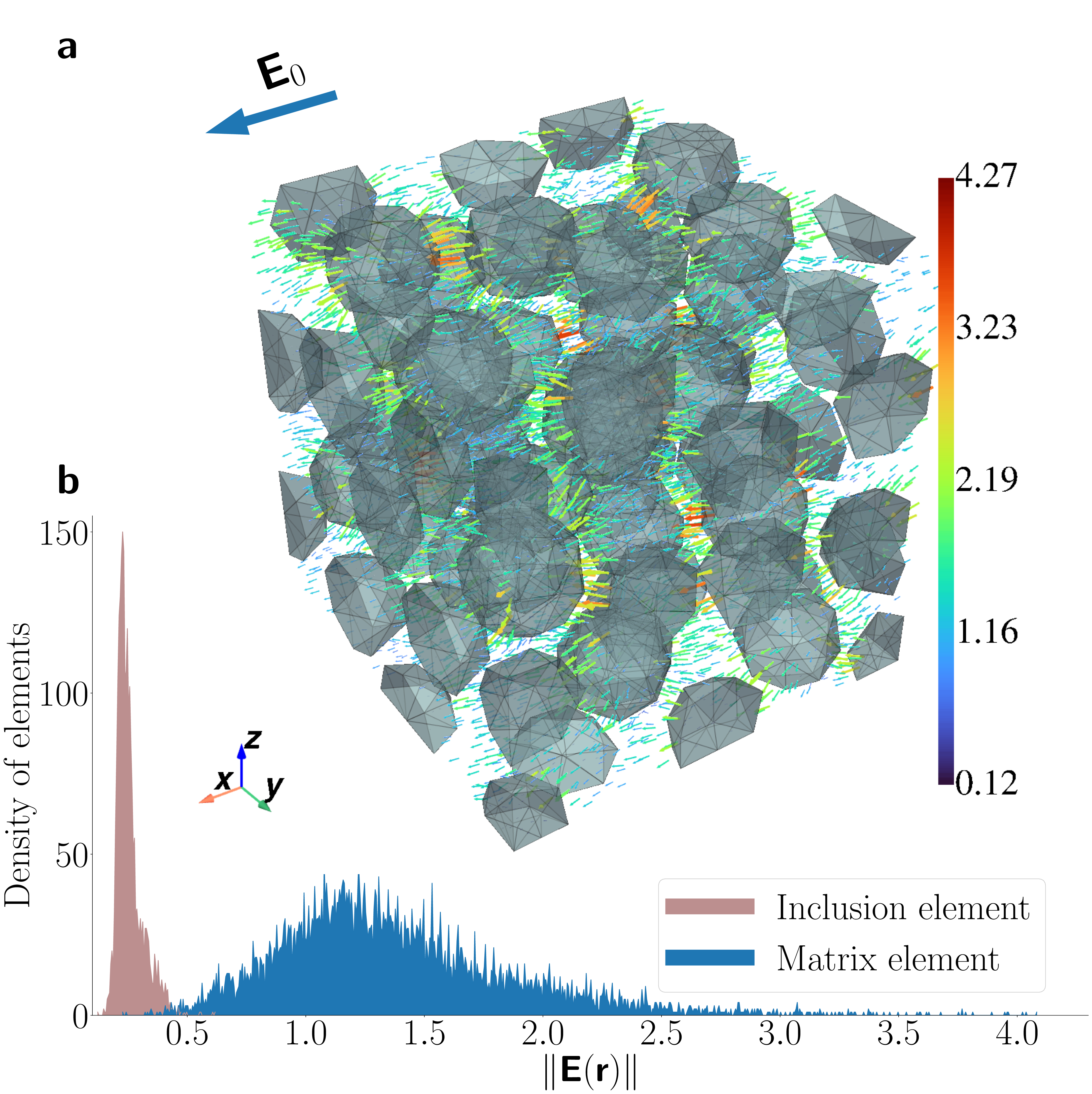}
\caption{\label{fig2}\textbf{Electric field distribution in the RVE.} \textbf{a}, Three-dimensional visualization
of the real component of the normalized electric field under $\mathbf{x}$-direction
excitation ($\mathbf{E}_{0}=\hat{\mathbf{x}}$). Arrows represent
element-averaged field vectors, with orientation showing field direction,
and length/color indicating magnitude. \textbf{b}, Statistical distribution
of field magnitudes across all elements.}
\end{figure}

Quantifying the depolarization field magnitude presents a challenge
for systems with multiple inclusions, as traditional methods only
yield rigorous solutions \cite{Choy2015} for isolated particles (e.g., spheres or
ellipsoids) in an effective medium. Here, we analyze the field distribution
under an applied field $\mathbf{E}_{0}$ (Fig.\ref{fig2}\textbf{a}). The field aligns
predominantly with the external direction, consistent with an isotropic
medium. Notably, the host matrix sustains a significantly stronger
field than the inclusions.

We quantify this effect by statistically analyzing the electric field
magnitude distribution across all elements in the RVE. Figure \ref{fig2}\textbf{b} plots
the element density as a function of field magnitude, demonstrating
that the average field in the host is 5.32 times as large as in the
inclusions. This provides direct evidence that depolarization dominates
the system\textquoteright s dielectric response.

\subsection{Origin of depolarization dominance}

While field distribution analysis confirms the dominance of depolarization
effects, the underlying mechanism requires examination of the $\bar{\epsilon}(q)\sim q$
relationship. Unlike the $E(\mathbf{k})\sim\mathbf{k}$ dispersion---where
$E(\mathbf{k})$ represents a real scalar quantity and Bloch states
are periodic in $\mathbf{k}$-space---$\bar{\epsilon}(q)$ constitutes
a complex dyadic tensor, with $q$ states lacking closed-form analytical
expressions.

\begin{figure}
\includegraphics[width=0.45\textwidth]{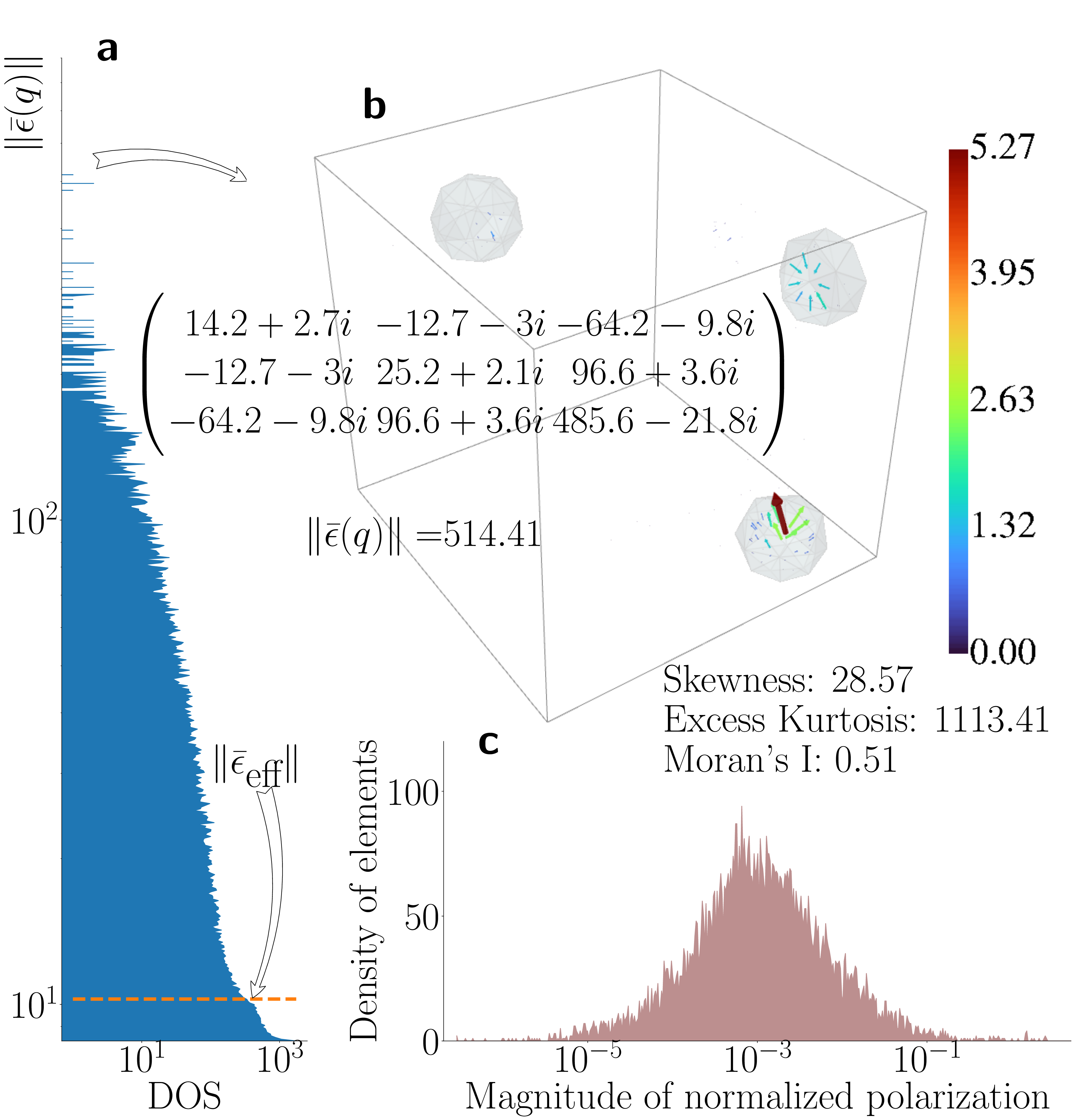}
\caption{\label{fig3} \textbf{Global analysis via the $\bar{\epsilon}(q)\sim q$ relationship.} \textbf{a,}
Density of states for $\|\bar{\epsilon}(q)\|$, with
the effective value $\|\bar{\epsilon}_{\mathrm{eff}}\|$ (10.25)
indicated (dashed line). \textbf{b,} Spatial distribution of the real component
of normalized electric polarization for the highest-$\|\bar{\epsilon}(q)\|$
state, showing localized field patterns (inset: corresponding permittivity
tensor and its norm). \textbf{c,} Statistical distribution of polarization
magnitudes for this state, characterized by extreme skewness (28.57),
kurtosis (1113.41), and spatial correlation (Moran's I = 0.51) as
shown in inset.}
\end{figure}

We adopt a two-stage analytical approach. First, we compute the density
of states for $\|\bar{\epsilon}(q)\|$ (Fig.\ref{fig3}\textbf{a}), revealing
a broad distribution spanning 8.01 to 514.11. The majority of states
cluster near $\|\bar{\epsilon}_{\mathrm{eff}}\|=10.25$,
while a small subset exhibits substantially larger norms that disproportionately
influence the system's dielectric response.

To understand these high-norm states, we examine their polarization
patterns. The highest-norm state (Fig.\ref{fig3}\textbf{b}) displays striking localization,
with polarization confined to discrete inclusions and their immediate
vicinity. Statistical analysis of the normalized polarization magnitude
(Fig.\ref{fig3}\textbf{c}) follows a logarithmic normal distribution \cite{Crow1988} spanning seven
orders of magnitude, characterized by extreme values that dominate
the spatial pattern.
We quantify this localization through three statistical measures:
\begin{itemize}
\item Skewness and excess kurtosis \cite{cramer1946mathematical} of polarization magnitude (indicating
extreme-value dominance).
\item Moran\textquoteright s I \cite{grekousis2020spatial} index (spatial clustering of extreme values;
range: 0--1).
\end{itemize}
The representative state in Fig.\ref{fig3}\textbf{b} exhibits values of 28.57, 1113.41,
and 0.51 for these respective metrics. 
All states exhibit similar localization (minimum values: 3.86, 17.9,
and 0.34, respectively), confirming that polarization remains confined
to discrete regions rather than propagating throughout the material.

This analysis reveals that dielectric behavior is governed by short-range
interactions, explaining the dominance of depolarization effects:
inclusions interact primarily with adjacent host material rather than
distant counterparts. A crucial remaining question concerns the precise
spatial decay of these interactions, which we address through local
analysis in the following section.

\subsection{Interaction length determination}

\begin{figure*}
\centering
\includegraphics[width=0.95\textwidth]{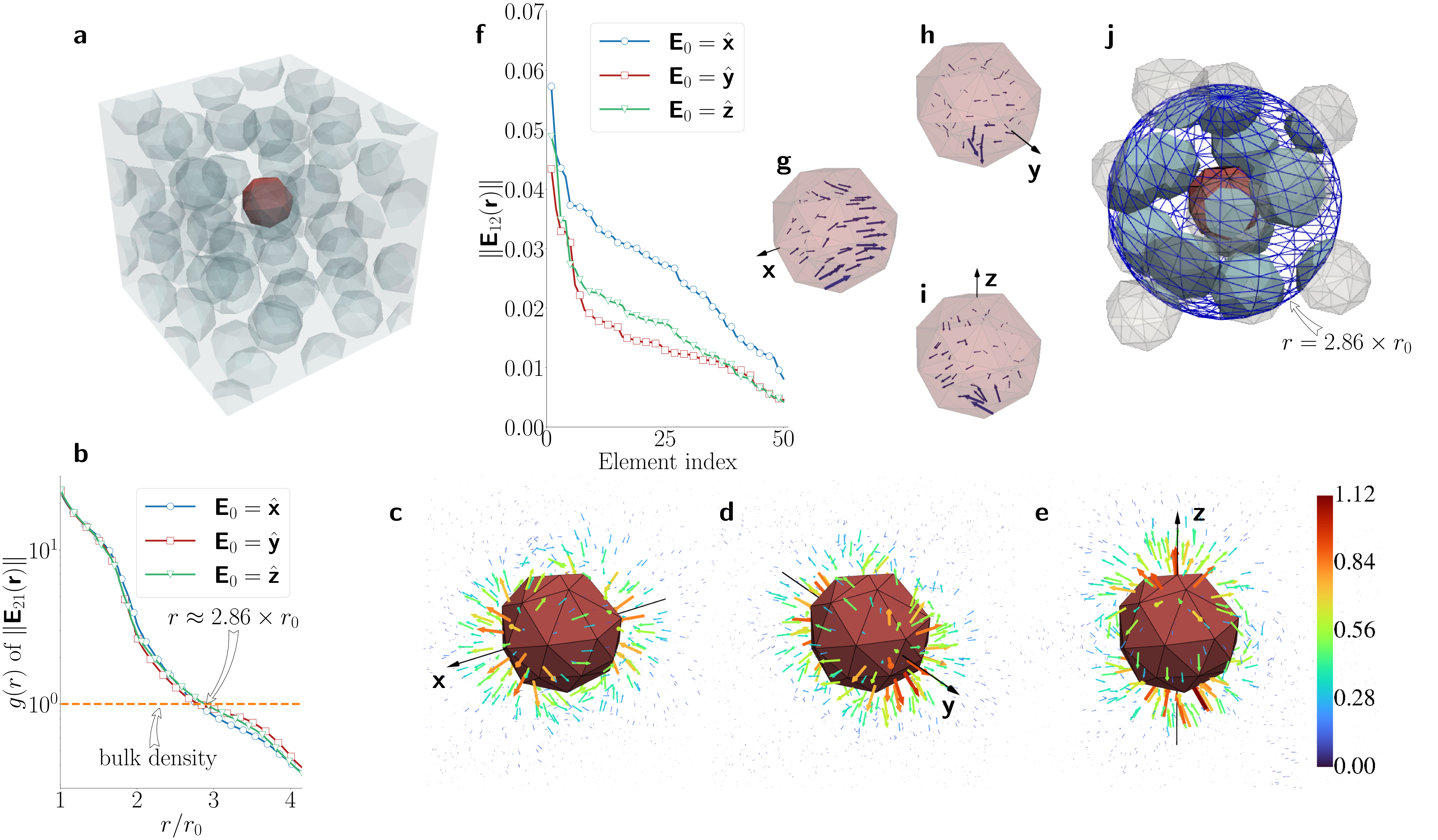}
\caption{\label{fig4} \textbf{Single inclusion response and local field characteristics.} \textbf{a,} Schematic
of the model showing the central inclusion (red, $\mathcal{H}_{1}$)
within the RVE. \textbf{b,} Radial correlation function $g(r)$ of $\|\mathbf{E}_{21}(\mathbf{r})\|$
for $\mathbf{x}$-, $\mathbf{y}$-, and $\mathbf{z}$-direction applied
fields, demonstrating consistent decay behavior. \textbf{c-e,} Three-dimensional
distributions of the real component of $\mathbf{E}_{21}(\mathbf{r})$
for $\mathbf{x}$ (\textbf{c}), $\mathbf{y}$ (\textbf{d}), and $\mathbf{z}$ (\textbf{e}) excitations,
showing dipole-like field patterns. \textbf{f}, Sorted magnitude distribution
of $\|\mathbf{E}_{12}(\mathbf{r})\|$ within $\mathcal{H}_{1}$
for different field directions. \textbf{g-i,} Corresponding real component
of $\mathbf{E}_{12}(\mathbf{r})$ distributions in the inclusion for
$\mathbf{x}$ (\textbf{g}), $\mathbf{y}$ (\textbf{h}), and $\mathbf{z}$ (\textbf{i}) excitations.
\textbf{j,} Local microstructure within the interaction range (2.86 inclusion
radii).}
\end{figure*}

To quantify the interaction range, we define $\mathcal{H}_{1}$ as
a single spherical inclusion near the RVE center (minimizing boundary
effects; Fig.\ref{fig4}\textbf{a}) and $\mathcal{H}_{2}$ as the remaining volume.
The field $\mathbf{E}_{21}(\mathbf{r})$ generated by $\mathcal{H}_{1}$
in its surroundings---computed via Eq.(\ref{eq:theory_la4}) for $\mathbf{E}_{0}$
aligned along $\mathbf{x}$, $\mathbf{y}$, and $\mathbf{z}$ directions
(Fig.\ref{fig4}\textbf{c-e})---reveals dipole-like patterns aligned with the applied
field. The rapid radial decay of these fields confirms short-range
interactions, consistent with our earlier conclusions.

We quantify the interaction length through the radial correlation
function $g(r)$ of the field magnitude (Fig.\ref{fig4}\textbf{b}). For all three field
orientations, $g(r)$ decays monotonically from the inclusion surface,
falling below unity (the average magnitude density) at about 2.86
inclusion radii. This critical distance---analogous to a Debye-like
screening length---defines the effective interaction range: a spherical
shell extending from the inclusion surface to 2.86 radii.

This limited range (insufficient to encompass adjacent inclusions)
demonstrates that interactions are confined to the immediate host
medium, corroborating our $q$-state analysis and confirming short-range
dominance in the dielectric response.

\subsection{Local structure-property relationship}

The interaction length defines both the spatial extent over which
an inclusion influences its surroundings and the boundary within which
neighboring structures modulate its dielectric response. This localized
interaction enables a quantitative structure-property analysis by
focusing solely on the relevant neighborhood around each inclusion.

The inclusion\textquoteright s effective dielectric response is given
by:

\begin{equation}
\bar{\epsilon}_{\mathrm{eff}}^{\mathrm{inc}}=\bar{\epsilon}_{\mathrm{eff}}+\frac{(\bar{\chi}_{11}+\bar{\chi}_{12})}{\phi_{\mathrm{inc}}},
\end{equation}
where $\phi_{\mathrm{inc}}=0.56\%$ is the inclusion\textquoteright s
volume fraction. The resulting dielectric tensor,

\begin{eqnarray*}
&&\bar{\epsilon}_{\mathrm{eff}}^{\mathrm{inc}}= \\
&&\left(\begin{array}{ccc}
15.249-0.756i & 0.269-0.019i & 0.061-0.005i\\
0.252-0.017i & 16.281-0.826i & -0.43+0.031i\\
0.004-0.0i & -0.336+0.023i & 15.848-0.798i
\end{array}\right)
\end{eqnarray*}
reveals two key features. First, the average permittivity (15.793\textminus 0.793j)
is only about one-third of the intrinsic inclusion value (50.0\textminus 5.0j),
confirming strong depolarization effects. Second, while the tensor
remains diagonally dominant, the 7\% variation among its diagonal
components---significantly exceeding the system\textquoteright s
global anisotropy (\ensuremath{\sim}1\%)---demonstrates how local
structural variations induce dielectric anisotropy at the inclusion
scale.
This local anisotropy is further evidenced by the orientation-dependent
variations in the perturbing field $\mathbf{E}_{12}(\mathbf{r})$
within the central inclusion (computed via Eq.(\ref{eq:theory_la5});
Fig.\ref{fig4}\textbf{g-i}). Statistical analysis (Fig.\ref{fig4}\textbf{f}) shows that the response
along the $\mathbf{x}$-direction exceeds other orientations by \ensuremath{\sim} 50\%,
underscoring the emergence of local anisotropy despite global isotropy.

To quantify the local microstructure within the interaction volume
(Fig.\ref{fig4}\textbf{j}), we introduce the fabric tensor $\bar{T}_{\mathrm{fab}}$
(adapted from trabecular bone studies to assess structural anisotropy) \cite{odgaard1997fabric}:

\begin{eqnarray*}
\bar{T}_{\mathrm{fab}}=\left(\begin{array}{ccc}
0.3430 & -0.0029 & -0.0014\\
-0.0029 & 0.3241 & 0.0019\\
-0.0014 & 0.0019 & 0.3329
\end{array}\right).
\end{eqnarray*}
Notably, $\bar{T}_{\mathrm{fab}}$ exhibits anisotropy comparable
to $\bar{\epsilon}_{\mathrm{eff}}^{\mathrm{inc}}$. A precise linear
relationship emerges between the two:

\begin{equation}
\mathrm{diag}(\bar{\epsilon}_{\mathrm{eff}}^{\mathrm{inc}})=-\beta diag(\bar{T}_{\mathrm{fab}})+\alpha\bar{I},
\end{equation}
with $\beta=(54.72\pm2.90)-(3.71\pm0.28)i$ and $\alpha=(34.03\pm0.97)-(2.03\pm0.09)i$. 

The negative correlation reflects depolarization: directions containing
more matrix material (higher $\bar{T}_{\mathrm{fab}}$ components)
exhibit stronger depolarization, reducing the effective permittivity
along those orientations. This quantitative relationship directly
links local microstructure to dielectric properties through fundamental
physical mechanisms.

\section{Conclusion}

In summary, we present a theoretical framework for classical many-body
interacting systems, recast in the language of operators, and demonstrate
its utility through the dielectric response of a random spherical
dispersion. Beyond predicting effective permittivity, our approach
reveals underlying mechanisms---globally via $q$-state decomposition
and locally through interaction-length analysis and structure-property
relationships. This framework not only extends traditional effective
medium theories but also provides a quantitative tool for probing
material properties governed by many-body interactions.

The implications are broad. Mathematically equivalent problems---magnetic
permeability, thermal conductivity, elastic properties, and diffusion
in porous media---can be analyzed with the same rigor. Moreover,
the framework\textquoteright s incorporation of full many-body interactions
makes it ideal for studying critical phenomena, including phase transitions,
percolation thresholds, and singularities in material response.

The mathematical tools introduced here---particularly SSVD for non-Hermitian
operators and the block operator formalism for coupling analysis---may
find wider applications. Since any finite square matrix is similar
to a complex symmetric matrix, this approach could extend to other
dissipative systems, classical or quantum. The method also opens avenues
for studying cross-coupling between material properties, such as electromagnetics
or multiphysics interactions.

By unifying microstructure, non-Hermitian interactions, and emergent
properties, this work advances our understanding of composite systems
and offers a blueprint for exploring structure-property relationships
across condensed matter physics and materials science.

\begin{acknowledgments}
This work was supported by the National Natural Science Foundation of China (Grant No. 52073076).
\end{acknowledgments}

\appendix

\section{Method}

\textbf{Random inclusion generation.}
Inclusions were positioned via random sequential addition \cite{widom1966random}, an
efficient algorithm for generating non-overlapping particle distributions.
Particles were placed iteratively using uniform random coordinates,
with each new inclusion retained only if its centroid maintained a
minimum separation of 2.2 radii from all existing particles. This ensured well-dispersed inclusions without clustering, consistent with MG\textquoteright  s single-particle approximation.

\textbf{Calculation of $\bar{\epsilon}_{\mathrm{eff}}$.}
We computed $\bar{\epsilon}_{\mathrm{eff}}$ self-consistently using
Eq.(\ref{eq:theory_ee15}). The initial permittivity value was derived
from the volume average of the local permittivity. In each iteration
cycle, we first constructed the matrix $\left[L_{mn}\right]$ based
on the $\bar{\epsilon}_{\mathrm{eff}}$ obtained from the previous
cycle. We then sequentially calculated the inverse of $\left[L_{mn}\right]$,
$\langle\bar{\chi}(\mathbf{r})\rangle$ and $\bar{\epsilon}_{\mathrm{eff}}$.
The matrix inversion was performed via the SSVD method, as described
in Eq.(\ref{eq:theory_ga2}). To minimize numerical errors between
cycles, we initialized $\bar{\epsilon}_{\mathrm{eff}}$ not with the
full tensor from the previous cycle, but with its isotropic average,
$\frac{1}{3}\mathrm{Tr}(\bar{\epsilon}_{\mathrm{eff}})\bar{I}$. Convergence
was assessed by monitoring the ratio ${\|\langle\bar{\chi}(\mathbf{r})\rangle\|}/{\|\bar{\epsilon}_{\mathrm{eff}}\|}$,
ensuring that the condition $\langle\bar{\chi}(\mathbf{r})\rangle=0$
was satisfied within the prescribed numerical tolerance.

\textbf{Generation of $\left[L_{mn}\right]$.}
The matrix $\left[L_{mn}\right]$ was constructed using H-SWG basis
functions, which are well-suited for modeling complex geometries  \cite{Zhang2015}.
These basis functions are defined on each face of a tetrahedral element
and extend into its volume as: 

\begin{equation}
\mathbf{f}_{n}(\mathbf{r}) =
\begin{cases}
\frac{a_{n}}{3\mathscr{V}_{n}}(\mathbf{r}-\mathbf{r}_{n}) & \mathbf{r}\in\mathscr{V}_{n},\\
0 & \text{otherwise},
\end{cases}
\end{equation}
where $a_{n}$ is the area of the face, $\mathbf{r}_{n}$ denotes
the coordinate of the tetrahedron vertex opposite the face, and $\mathscr{V}_{n}$
is the volume of the element.

For the Green\textquoteright s function, we adopted the form for an
isotropic medium \cite{Weiglhofer1993}, justified by the uniform distribution of inclusions
and the macroscopic isotropy of the effective medium: 
\begin{equation}
\bar{G}_{ee}(\mathbf{r},\mathbf{r'})=i\omega\mu_{0}\left(\bar{I}+\frac{\nabla\nabla}{k^{2}}\right)\frac{\exp(ik|\mathbf{r}-\mathbf{r'}|)}{4\pi|\mathbf{r}-\mathbf{r'}|},
\end{equation}
where $\mu_{0}$ is the vacuum magnetic permeability (assuming a non-magnetic
medium), and $k=\omega\sqrt{\mu_{0}\bar{\epsilon}_{\mathrm{eff}}}$
is the wavevector magnitude in the effective medium. This introduces
a dependence of $\left[L_{mn}\right]$ on $\bar{\epsilon}_{\mathrm{eff}}$,
necessitating the self-consistent calculation described earlier.

Following the testing procedure in Eqs.(\ref{eq:theory_ee6}--\ref{eq:theory_ee8}),
$\left[L_{mn}\right]$ was generated with dimensions $N\times N$,
where $N$ corresponds to the total number of basis functions.

\textbf{Numerical implementation.}
Numerical computations were performed on a high-performance computing
system equipped with dual Intel Xeon Platinum 8383C processors and
two NVIDIA V100s GPUs, supported by 1 TB of system memory. The custom
C implementation leveraged Intel Math Kernel Library \cite{intelmkl} optimizations
for CPU computations and CUDA \cite{nvidiacuda} acceleration for GPU operations, with
OpenMP parallelization employed throughout to enhance computational
efficiency.

For the SSVD routine, we developed a custom implementation as existing
numerical libraries lack support for this specialized operation. Our
implementation follows established numerical algorithms \cite{bunsegerstner1988singular, qiao2001fast} beginning
with matrix reduction to tridiagonal form using a stabilized Lanczos
approach, followed by diagonalization through an optimized divide-and-conquer
method. Numerical accuracy was verified by monitoring two error measures:
the orthogonality error

\begin{equation}
\mathrm{Err_{orth}}=\frac{\| QQ^{H}-I\|}{\| I\|},
\end{equation}
and the overall reconstruction error 
\begin{equation}
\mathrm{Err_{ssvd}}=\frac{\| Q\varLambda Q^{T}-\left[L_{mn}\right]\|}{\|\left[L_{mn}\right]\|}.
\end{equation}

\textbf{Effective permittivity calculated by MG theory.}
For a three-dimensional, two-phase composite system comprising inclusions
(permittivity $\epsilon_{\mathrm{i}}$) dispersed in a continuous
matrix (permittivity $\epsilon_{\mathrm{m}}$), the effective permittivity
$\epsilon_{\mathrm{MG}}$ can be derived using the MG
effective medium approximation. The formulation accounts for the volume
fraction of inclusions ($\phi_{\mathrm{i}}$) and is given by \cite{maxwell1904colours}:

\begin{equation}
\frac{\epsilon_{\mathrm{MG}}-\epsilon_{\mathrm{i}}}{\epsilon_{\mathrm{MG}}+2\epsilon_{\mathrm{m}}}=\phi_{\mathrm{i}}\frac{\epsilon_{\mathrm{m}}-\epsilon_{\mathrm{i}}}{\epsilon_{\mathrm{m}}+2\epsilon_{\mathrm{i}}}.
\end{equation}
This expression assumes dipolar interactions dominate and neglects
higher-order multipole effects, providing a closed-form approximation
for dilute systems with low inclusion concentration.

\textbf{Visualization of the field distribution.}
The field distribution is visualized element-wise by averaging over
the four H-SWG basis functions defined on each element:

\begin{equation}
\mathbf{E}_{j}^{\mathrm{ele}}=\frac{1}{\chi_{j}}\sum_{t=1}^{4}e_{t}\frac{1}{\mathscr{V}_{j}}\int_{\mathscr{V}_{j}}\mathbf{f}_{t}(\mathbf{r})\,dv,
\end{equation}
where $\mathbf{E}_{j}^{\mathrm{ele}}$ is a 3D complex vector representing
element $j$, enabling straightforward visualization of the field
distribution in complex 3D models.

Since complex vectors cannot be fully represented by a single spatial
arrow (due to their six independent components), we visualize their
real and imaginary parts separately \cite{lindell1996methods}:

\begin{equation}
\mathbf{E}_{j}^{\mathrm{ele}}=\mathbf{f}_{\mathrm{re}}+\mathbf{f}_{\mathrm{im}}\,i.
\end{equation}
Here, $\mathbf{f}_{\mathrm{re}}$ and $\mathbf{f}_{\mathrm{im}}$
are real vectors corresponding to the physical time-harmonic field
$\mathbf{F}(t)$:

\begin{equation}
\mathbf{F}(t)=\mathrm{Re}\left\{ \mathbf{E}_{j}^{\mathrm{ele}}\,e^{i\omega t}\right\} =\mathbf{f}_{\mathrm{re}}\cos(\omega t)-\mathbf{f}_{\mathrm{im}}\sin(\omega t),
\end{equation}
where $\omega$ is the angular frequency and $t$ is time. The vectors
$\mathbf{f}_{\mathrm{re}}=\mathbf{F}(0)$ and $\mathbf{f}_{\mathrm{im}}=\mathbf{F}(\pi/2\omega)$
represent snapshots of the field at $t=0$ and $t=\pi/2\omega$, respectively.
The full time-dependent field is a linear combination of these components.
Plotting both $\mathbf{f}_{\mathrm{re}}$ and $\mathbf{f}_{\mathrm{im}}$
thus provides a complete representation of the field dynamics. All
visualizations were generated using PyVista \cite{sullivan2019pyvista} with a custom Python script.

\textbf{Visualization of the normalized polarization distribution.}
The normalized polarization distribution for state $q$ is visualized
similarly, with the coefficients $\{e_{n}\}$ replaced by the singular
vector $q_{n}$ and the electric field replaced by the normalized
polarization:

\begin{equation}
\mathbf{p}_{j}^{\mathrm{ele}}=\sum_{t=1}^{4}(q_{n})_{t}\frac{1}{\mathscr{V}_{\mathrm{ave}}}\int_{\mathscr{V}_{j}}\mathbf{f}_{t}(\mathbf{r})\,dv,
\end{equation}
where $\mathbf{p}_{j}^{\mathrm{ele}}$ is the normalized polarization
for element $j$, and $(q_{n})_{t}$ is the $t$-th component of the
singular vector $q_{n}$. 

\textbf{Calculation of skewness, excess kurtosis and Moran's I.}
For the complex field vector $\mathbf{E}_{j}^{\mathrm{ele}}$, we
calculate its magnitude $M_{j}=\|\mathbf{E}_{j}^{\mathrm{ele}}\|$
across all elements ($N_{e}$ denotes the total number of elements).
The skewness of the magnitude distribution $\{M_{j}\}_{j=1}^{N_{e}}$
is given by \cite{cramer1946mathematical}:

\begin{equation}
\mathrm{skewness}=\frac{\sum_{j=1}^{N_{e}}(M_{j}-\bar{M})^{3}/N_{e}}{\sigma^{3}},
\end{equation}
where $\bar{M}=\frac{1}{N_{e}}\sum_{j=1}^{N_{e}}M_{j}$ is the mean
magnitude and $\sigma=\sqrt{\frac{1}{N_{e}}\sum_{j=1}^{N_{e}}(M_{j}-\bar{M})^{2}}$
is the standard deviation.

The excess kurtosis is calculated as \cite{cramer1946mathematical}:

\begin{equation}
\mathrm{excess\ kurtosis}=\frac{\sum_{j=1}^{N_{e}}(M_{j}-\bar{M})^{4}/N_{e}}{\sigma^{4}}-3.
\end{equation}

To compute Moran's I spatial autocorrelation metric, we first binarize
the magnitude distribution:

\begin{equation}
Y_{j}=\begin{cases}
1 & M_{j}\geq\bar{M}\\
0 & M_{j}<\bar{M}
\end{cases}
\end{equation}
This transformation focuses on spatial clustering of extreme values
while mitigating bias from the heavy-tailed magnitude distribution.
Moran's I is then defined as \cite{grekousis2020spatial}:

\begin{equation}
\mathrm{Moran's\ I}=\left(\frac{N_{e}}{\sum_{i,j}w_{ij}}\right)\left(\frac{\sum_{i,j}w_{ij}(Y_{i}-\bar{Y})(Y_{j}-\bar{Y})}{\sum_{i=1}^{N_{e}}(Y_{i}-\bar{Y})^{2}}\right),
\end{equation}
where $\bar{Y}=\frac{1}{N_{e}}\sum_{j=1}^{N_{e}}Y_{j}$ and the spatial
weight matrix $w_{ij}$ is defined as:

\begin{equation}
w_{ij}=\begin{cases}
1 & \text{if elements \ensuremath{i} and \ensuremath{j} share a tetrahedral face}\\
0 & \text{otherwise}
\end{cases}
\end{equation}

\textbf{Calculation of radial correlation function $g(r)$.}
We define the radial correlation function $g(r)$ \cite{hansen2013theory} of the field magnitude
$\|\mathbf{E}_{21}(\mathbf{r})\|$ as:

\begin{equation}
g(r)=\frac{\langle\rho(r)\rangle}{\rho_{0}},
\end{equation}
where $\langle\rho(r)\rangle$ represents the average field magnitude
density at distance $r$ from the inclusion centroid, and $\rho_{0}$
denotes the bulk field magnitude density.

The theoretical expression for $\langle\rho(r)\rangle$ is given by
the surface integral:

\begin{equation}
\langle\rho(r)\rangle=\frac{1}{4\pi r^{2}}\int_{S}\|\mathbf{E}_{21}(\mathbf{r})\|\,ds,
\end{equation}
which we implement numerically through element-wise interpolation:

\begin{equation}
\langle\rho(r)\rangle=\frac{1}{4\pi r^{2}}\sum_{i}S_{i}\cdot M_{i}.
\end{equation}
Here, $S_{i}$ is the intersection area between the spherical shell
at radius $r$ and element $i$, while $M_{i}$ is the element's average
field magnitude.

The bulk density $\rho_{0}$ is computed theoretically as:

\begin{equation}
\rho_{0}=\frac{1}{\mathscr{V}}\int_{\mathscr{V}}\|\mathbf{E}_{21}(\mathbf{r})\|\,dv,
\end{equation}
with its discrete counterpart:

\begin{equation}
\rho_{0}=\frac{1}{\mathscr{V}}\sum_{i}\mathscr{V}_{i}\cdot M_{i},
\end{equation}
where $\mathscr{V}_{i}$ is the volume of element $i$ and $\mathscr{V}$
is the volume of the RVE. 

\textbf{Calculation of the fabric tensor.}
We compute the fabric tensor using the star length distribution approach \cite{odgaard1997fabric}.
For a given inclusion centroid, the fabric tensor $\bar{T}_{\mathrm{fab}}$
is constructed from $N_{t}$ intercept lengths and their associated
orientations:

\begin{equation}
\bar{T}_{\mathrm{fab}}=\sum_{i=1}^{N_{t}}l_{i}(\mathbf{\hat{v}}_{i})\mathbf{\hat{v}}_{i}\mathbf{\hat{v}}_{i}^{T},
\end{equation}
where $\mathbf{\hat{v}}_{i}$ is a unit vector orientation, $\mathbf{\hat{v}}_{i}\mathbf{\hat{v}}_{i}^{T}$
forms the orientation dyad, and $l_{i}(\mathbf{\hat{v}}_{i})$ represents
the intercept length along direction $\mathbf{\hat{v}}_{i}$. Each
intercept segment passes through the inclusion centroid and terminates
at both ends by intersecting either adjacent inclusions or the interaction
length boundary.

To ensure unbiased sampling of the local structure, we generate uniformly
distributed unit vectors $\mathbf{\hat{v}}_{i}$ by normalizing random
Gaussian variates \cite{muller1959note}:

\begin{equation}
\mathbf{\hat{v}}_{i}=\frac{(x_{i},y_{i},z_{i})}{\sqrt{x_{i}^{2}+y_{i}^{2}+z_{i}^{2}}},
\end{equation}
where $\{x_{i}\}$, $\{y_{i}\}$, and $\{z_{i}\}$ are independent
Gaussian-distributed random numbers. We use $N_{t}=20,\!000$ sampling
directions to achieve sufficient angular uniformity in the representation. 

\textbf{Linear fit of effective permittivity versus fabric tensor.}
We perform separate linear regressions \cite{press2007numerical} for the real and imaginary
components of the complex-valued $\mathrm{diag}(\bar{\epsilon}_{\mathrm{eff}}^{\mathrm{inc}})$
against the real-valued $\mathrm{diag}(\bar{T}_{\mathrm{fab}})$.
For each component, we model the relationship as $y=\alpha+\beta x$,
where:
\begin{itemize}
\item For the real part: $y_{i}=\mathrm{Re}\{\mathrm{diag}(\bar{\epsilon}_{\mathrm{eff},i}^{\mathrm{inc}})\}$
and $x_{i}=\mathrm{diag}(\bar{T}_{\mathrm{fab},i})$ 
\item For the imaginary part: $y_{i}=\mathrm{Im}\{\mathrm{diag}(\bar{\epsilon}_{\mathrm{eff},i}^{\mathrm{inc}})\}$ 
\end{itemize}
The regression parameters $\mathbf{\theta}=(\alpha,\beta)^{\mathrm{T}}$
are determined via:

\begin{equation}
\mathbf{\theta}=(\mathbf{X}^{\mathrm{T}}\mathbf{\Sigma}^{-1}\mathbf{X})^{-1}\mathbf{X}^{\mathrm{T}}\mathbf{\Sigma}^{-1}\mathbf{y}
\end{equation}
where $\mathbf{X}$ is the design matrix with rows $(1,x_{i})$, and
$\mathbf{\Sigma}$ is the covariance matrix accounting for measurement
uncertainties and correlations. Parameter uncertainties are obtained
from the covariance matrix:

\begin{equation}
\mathrm{Cov}(\mathbf{\theta})=(\mathbf{X}^{\mathrm{T}}\mathbf{\Sigma}^{-1}\mathbf{X})^{-1}
\end{equation}
This yields the slope $\beta\pm\sigma_{\beta}$ and intercept $\alpha\pm\sigma_{\alpha}$,
where $\sigma_{\beta}=\sqrt{\mathrm{Cov}(\mathbf{\theta})_{22}}$
and $\sigma_{\alpha}=\sqrt{\mathrm{Cov}(\mathbf{\theta})_{11}}$.
The same procedure is applied independently to both the real and imaginary
components.

\end{document}